\documentclass{article}
\usepackage{graphicx}
\usepackage[small]{subfigure}
\usepackage{booktabs}
\usepackage{amsmath,amssymb,bm}
\usepackage{multirow,threeparttable}
\usepackage[hypertexnames=false]{hyperref}

\usepackage[accepted]{mlsys2025}
\makeatletter
\renewcommand{\Notice@String}{Preprint.}
\makeatother
\hypersetup{pdfauthor={Botao Zhu, Xiaoyi Fan, Yifei Zhu},pdfsubject={Preprint}}
\newcommand{\Call}[2]{\textsc{#1}(#2)}
\usepackage{xurl}
\mlsystitlerunning{Presage: Server-Aware Configuration and Predictive Scheduling}
\makeatletter
\AtBeginDocument{\let\normalsize\@normalsize\normalsize}
\makeatother

\begin{document}
\twocolumn[
\mlsystitle{Presage: Server-Aware Configuration and Predictive Scheduling for LLM Agent Serving}
\begin{mlsysauthorlist}
\mlsysauthor{Botao Zhu}{sjtu}
\mlsysauthor{Xiaoyi Fan}{jx}
\mlsysauthor{Yifei Zhu}{sjtu}
\end{mlsysauthorlist}
\mlsysaffiliation{sjtu}{Global College, Shanghai Jiao Tong University}
\mlsysaffiliation{jx}{JX Intelligence Technology Co., Ltd}
\mlsyscorrespondingauthor{Yifei Zhu}{yifei.zhu@sjtu.edu.cn}
\begin{center}
\texttt{zhubotao@sjtu.edu.cn}\quad
\texttt{xiaoyi.fan@ieee.org}\quad
\texttt{yifei.zhu@sjtu.edu.cn}
\end{center}
\mlsyskeywords{Large Language Models, LLM Agents, Resource Scheduling}
\vskip 0.3in
\begin{abstract}
LLM agents execute queries as workflows of dependent inference and tool stages, with different agent and model choices yielding different answer quality and resource demands. The latency of a configuration changes with server load and model residency, while uncertain stage durations and stage dependencies obscure which agents will finish soon and when each model will be needed. We present Presage, an end-to-end serving system that addresses these challenges through server-aware adaptation and predictive scheduling. Its adapter combines a graph-based, query-specific quality predictor with a server-aware latency estimator and efficient Pareto sampling to select configurations predicted to meet quality targets at low latency. During execution, Presage exploits correlations among stages to predict each agent’s remaining service time and prioritize execution, and forecasts when stage dependencies will release model demand to guide GPU allocation and model switching. Across diverse workloads, Presage reduces the average end-to-end latency by 10.4–97.1\% while improving answer quality by 3.1–80.3\% over existing methods.
\end{abstract}
]
\printAffiliationsAndNotice{}
\section{Introduction}

Large language model (LLM) agents are increasingly used to solve complex tasks that a single-pass LLM inference cannot reliably handle. LLM agents organize multiple LLM calls, tool invocations, code execution, and result aggregation into structured workflows, enabling broader capabilities and richer reasoning patterns. As a result, LLM agents have shown strong potential in coding, mathematical reasoning, reading comprehension, and other tool-intensive workloads. Many agents have been proposed to handle heterogeneous tasks, including human-crafted agents such as ReAct \cite{react}, Self-Consistency \cite{selfconsistency}, and Debate \cite{llmdebate}, and automatically discovered agents by GPTSwarm \cite{gptswarm} and AFlow \cite{aflow}.

{LLM agents exhibit two key characteristics. First, they demonstrate substantial performance heterogeneity across tasks, queries, and underlying LLMs. Different agents embody distinct reasoning strategies, while different LLMs vary in their ability to support these strategies. Consequently, no single agent-model configuration, i.e., the selected agent and the model(s) assigned to its inference stages, is uniformly optimal \cite{maas,masrouter}. Our empirical study shows that different configurations can lead to up to a $1.8\times$ difference in answer quality and a $5\times$ difference in end-to-end (E2E) service time, motivating recent efforts on \textit{agent adaptation} \cite{maas,aflow} that adaptively choose configurations per task and query. Second, LLM agents are multi-stage workflow graphs with highly dynamic execution: even for the same agent, different queries may trigger different execution paths, prompt/completion lengths, and tool invocations. For example, iterative code-generation workflows can range from 3 to 15 steps across queries \cite{llmsched}, leading to highly variable resource demands and service times.}

These two characteristics make low-latency, high-quality agent serving challenging. First, at the adaptation level, the system searches a large agent-model configuration space with a non-trivial configuration-performance relationship, further complicated by dynamic server state (model residency, workload intensity, and cache availability) that existing methods ignore yet substantially affects E2E latency. Second, at the scheduling level, uncertain agent execution makes resource demands and execution topology hard to estimate a priori, while the topology-dependent model demand of multi-model agents is shaped by stage dependencies; reactively scheduling models based only on currently pending requests can cause mistimed model switches, poor GPU allocation, and unnecessary loading overhead: a model with little ready work may soon become a bottleneck once upstream stages finish, whereas a model with high aggregate demand may still be a poor switching target if most of its dependent stages remain blocked.

{Existing methods exhibit significant inefficiencies. Adaptation methods \cite{aflow,maas,masrouter} mainly optimize answer quality or token cost while ignoring real-time server state, making static and suboptimal decisions such as choosing non-resident models or persisting with compute-intensive agents under overload. Meanwhile, existing serving systems \cite{vllm,parrot,ayo} target single-request serving or static LLM workflows and therefore fall short in adaptive agent serving: at the agent level, they fail to handle execution uncertainty, causing severe head-of-line (HoL) blocking; at the model level, they either lack multi-model coordination or operate at the request level \cite{infaas,awssagemaker,cocktail}, agnostic to topology-dependent model demand, triggering frequent model switches and myopic decisions that substantially degrade E2E latency.}

\begin{figure}[t]

  \centering
    \includegraphics[width=0.97\linewidth]{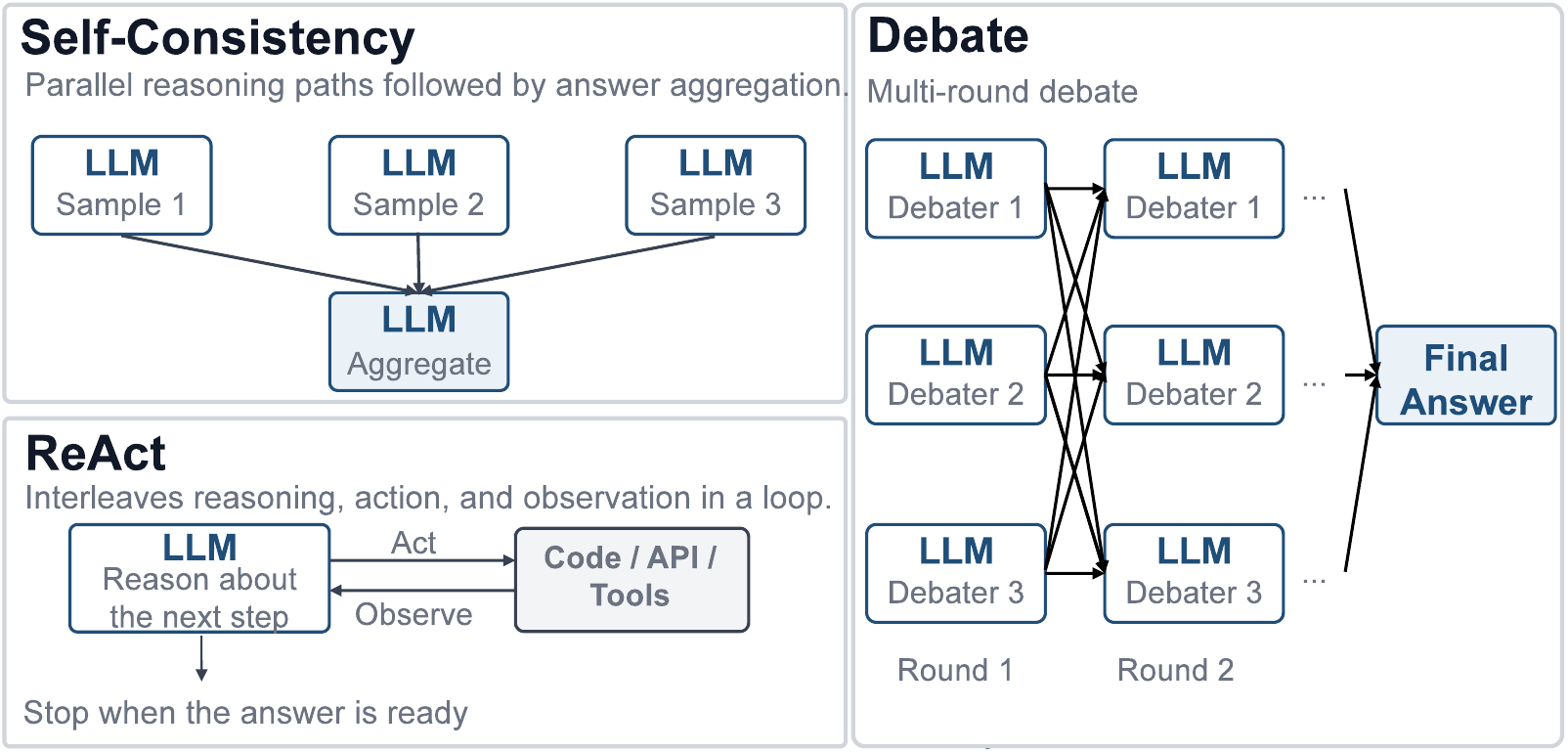}
            \caption{Examples of LLM agents.}
        \label{fig:agentexample}
\end{figure}

{To this end, we present Presage, an end-to-end agent serving system with server-aware adaptation and predictive scheduling. In the adaptation layer, Presage uses a tailored HGNN-based predictor (\S\ref{sec:hgnn_predictor}) to estimate query-specific answer quality and a lightweight server-aware latency estimator (\S\ref{sec:server_aware_latency}) that captures key-value (KV) cache reuse, model residency, and workload intensity; the adapter then efficiently searches the agent-model space to select a configuration that satisfies the quality target with minimal E2E latency (\S\ref{sec:pareto_sampling}). In the scheduling layer, Presage performs predictive scheduling: a probabilistic agent scheduler (\S\ref{sec:agentscheduler}) exploits intra-agent spatial and temporal correlations to continuously predict each agent's remaining service time and prioritizes short-running agents to reduce HoL blocking, while a topology-aware model scheduler (\S\ref{sec:modelscheduler}) forecasts topology-dependent future demand and timely allocates GPU resources, avoiding unnecessary model switching and resource waste.}

{We implement Presage in Python with approximately 5K lines of code and evaluate it on diverse agentic workloads constructed from the Azure inference trace and popular LLM benchmarks covering math reasoning, code generation, and reading comprehension tasks, using six popular LLMs and nine agent candidates. Comprehensive experiments demonstrate that Presage reduces average JCT by up to 97.1\% and P95 latency by up to 91.8\% compared to state-of-the-art adaptation methods and serving systems, while improving answer quality by up to 80.3\% with negligible online overhead.}

Our main contributions are summarized as follows:
\begin{itemize}
\item We identify limitations of existing methods in serving LLM agents, i.e., server-agnostic agent adaptation, coarse-grained agent scheduling, and lack of multi-model coordination.

\item We design Presage, an end-to-end agent serving framework that combines a server-aware adaptation layer and {a predictive scheduling layer}.

\item We implement Presage and evaluate it on diverse agentic workloads, showing that it substantially improves serving efficiency while maintaining high answer quality.
\end{itemize}
\section{Background}
\label{sec:background}

\subsection{LLM Inference}
Modern LLMs are commonly built on decoder-only Transformer \cite{transformer} architectures, such as Llama, Qwen, and GPT \cite{llama, qwen, gpt4}, where stacked self-attention and feed-forward layers autoregressively generate tokens conditioned on previous context. Serving an LLM request typically consists of two phases. In the prefill phase, the model processes the entire input prompt in parallel and computes the key-value (KV) states for all prompt tokens. In the decode phase, the model generates completion tokens sequentially, where each new token depends on the previously generated context. To avoid repeatedly recomputing past attention states and accelerate inference, it is a common practice to maintain a key-value (KV) cache across decoding steps \cite{orca, vllm}.

\subsection{LLM Agent}
\label{sec:bg_agent}

An LLM agent refers to a structured execution graph composed of multiple LLM invocations and auxiliary operations (e.g., tool calls, code execution, or result aggregation) that collaboratively solve a user query. Each agent is defined by a workflow graph $G = (V,E)$, where $V$ is a set of executable stages, and $E$ is the set of inter-stage dependencies. Each stage $v$ either runs LLM inference with a specific model or invokes a tool, and  passes the output to its successor stages.

{Early agents were largely human-designed, e.g., ReAct \cite{react}, Self-Consistency \cite{selfconsistency}, and Debate \cite{llmdebate} (Fig. \ref{fig:agentexample}). They outperform single-pass inference on reasoning- and tool-intensive tasks, but their fixed, manually tuned structures limit adaptability across diverse tasks and query difficulties. To overcome this, methods like AFlow \cite{aflow} and MaAS \cite{maas} automatically discover task- or query-adaptive agents by searching over compositions of empirically effective primitives, producing agents with diverse structures and execution patterns that substantially improve LLMs' performance.}

\section{Motivation}
\label{sec:motivation}

{We conduct motivating experiments on two H800-96GB GPUs, each split into two 48GB slices via NVIDIA MIG \cite{mig} to serve four LLMs simultaneously with vLLM \cite{vllm}. User queries are constructed from HotpotQA \cite{hotpotqa} (reading comprehension), HumanEval \cite{humaneval} (code generation), and MATH \cite{mathdataset} (math reasoning), with real-world arrivals synthesized from the Azure Inference Trace \cite{llmtrace}. We use six popular LLMs from different providers: Ministral-3-8B, Llama-3.2-3B, Llama-3.1-8B, Qwen3-0.6B, Qwen3-4B, and Phi-4-14B. Motivated by on-demand model loading in Amazon SageMaker \cite{awssagemaker}, we implement a demand-driven model scheduler that loads the model with the most inference requests whenever a GPU is idle, and a First Come, First Served (FCFS) agent scheduler. The following observations use this setting unless otherwise specified.}

\subsection{Motivation for Server-aware Adaptation}
\label{sec:motivation_serveraware}

\begin{figure}[t]
    \centering

    \subfigure[Answer quality across benchmarks.]{\includegraphics[width=0.48\columnwidth]{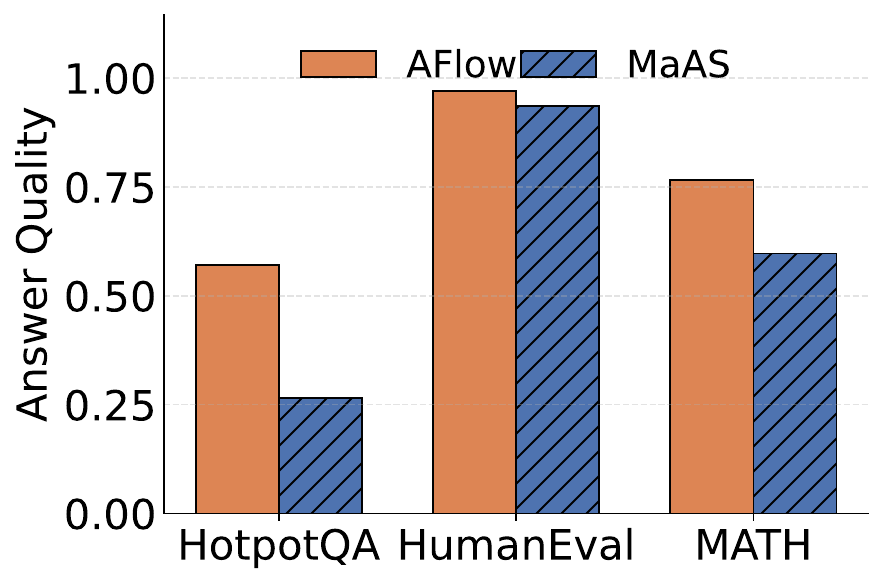}\label{fig:trace_acc}}
    \hfill
    \subfigure[Average JCT.]{\includegraphics[width=0.48\columnwidth]{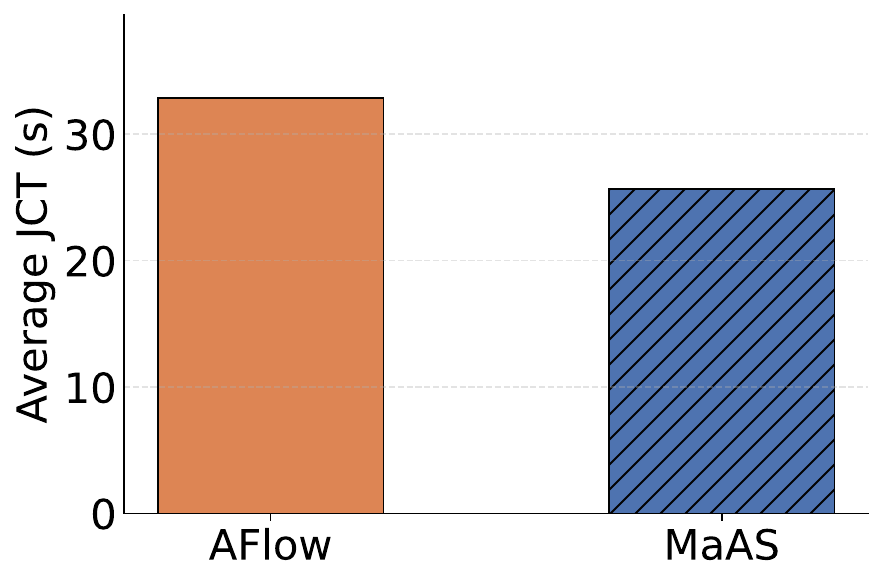}\label{fig:trace_token}}

    \subfigure[Overall token usage.]{\includegraphics[width=0.48\columnwidth]{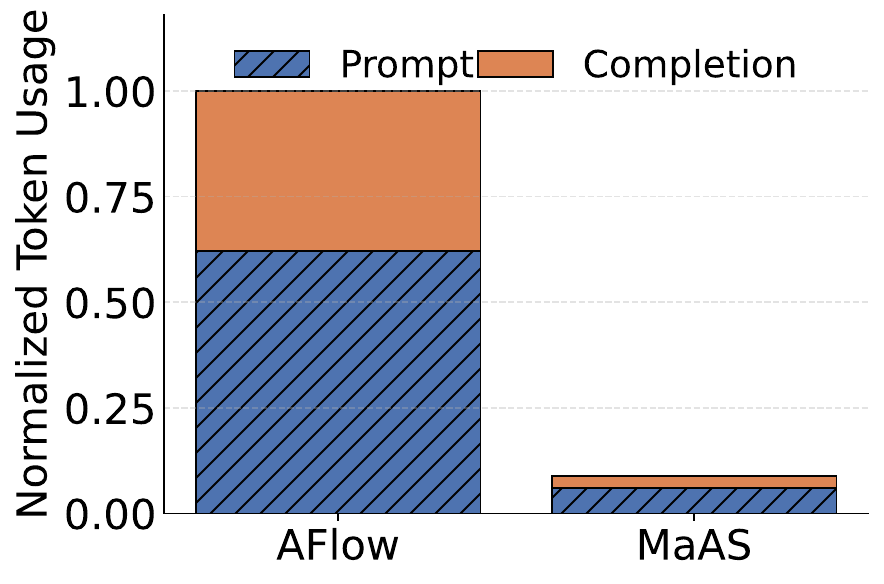}\label{fig:trace_jct}}
    \hfill
    \subfigure[Runtime breakdown.]{\includegraphics[width=0.48\columnwidth]{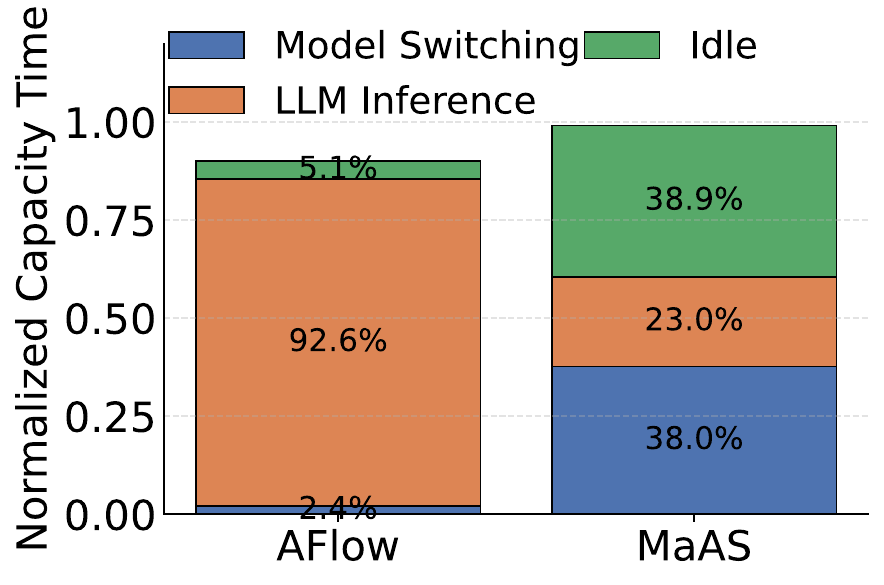}\label{fig:trace_breakdown}}

    \caption{Inefficiency analysis of AFlow and MaAS on (a) average answer quality, (b) average JCT, (c) overall token usage and (d) runtime breakdown.}
    \label{fig:motivation1}
\end{figure}

\textbf{Observation 1: Existing agent adaptation methods are highly inefficient due to mismatched objectives and the lack of server awareness.}
{Existing adaptation methods optimize objectives such as answer quality or token cost based purely on the agent itself. Since server state, especially model residency and workload, directly affects execution latency, ignoring it yields agents that are locally optimal at the agent level yet incur substantially higher end-to-end job completion time (JCT) at the serving engine.}

{We examine two SOTA adaptation methods, AFlow \cite{aflow} and MaAS \cite{maas}, in Fig. \ref{fig:motivation1}. AFlow achieves the highest answer quality on all three benchmarks but incurs a high average JCT of 32.86s, as it repeatedly selects expensive agents regardless of server load. MaAS reduces total token usage to less than one-tenth of AFlow yet only lowers average JCT to 25.68s: it often selects cheap but non-resident models, so model switching accounts for 38.0\% of total GPU capacity time while another 38.9\% remains idle. Worse, this token-driven adaptation reduces the answer quality on HotpotQA by more than half.}

\begin{figure}[t]
  \centering

  \begin{minipage}[t]{0.48\linewidth}
    \centering
    \includegraphics[width=\linewidth]{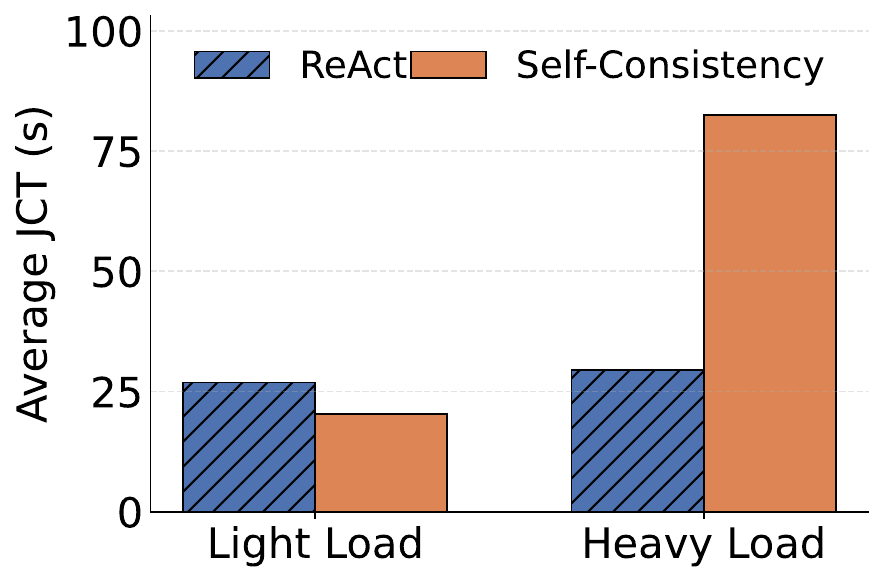}
    \caption{Average JCT under light and heavy load using different agents.}
    \label{fig:agentcompare}
  \end{minipage}
  \hfill
  \begin{minipage}[t]{0.48\linewidth}
    \centering
    \includegraphics[width=\linewidth]{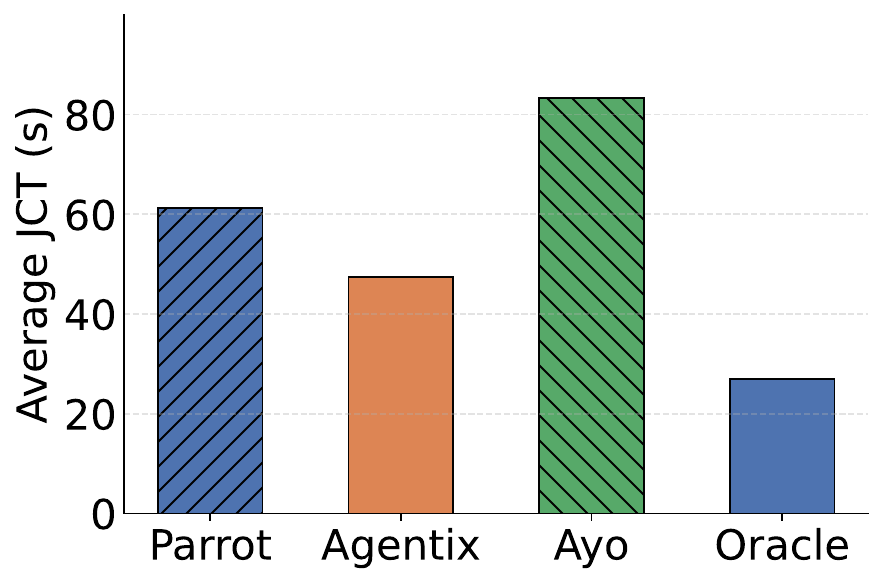}
    \caption{JCT comparison among four agent schedulers.}
    \label{fig:motivation3}
  \end{minipage}

\end{figure}

\textbf{Observation 2: Agent-level optimality breaks down without server-aware execution context.}
The latency impact of agent complexity is not intrinsic to the agent itself; rather, it strongly depends on the runtime server state. To demonstrate this point, we compare two agents, ReAct and Self-Consistency, using the Ministral-3-8B model and the MATH \cite{mathdataset} dataset. Their workflows are shown in Fig. \ref{fig:agentexample}: Self-Consistency merges answers from multiple samples, while ReAct performs iterative reasoning and action. We construct two workloads by scaling the same trace: a light-load setting and a heavy-load setting with a 3$\times$ arrival rate.

The two agents achieve nearly identical answer quality in both settings (62.1\% for ReAct and 61.2\% for Self-Consistency). Either agent appears reasonable if one considers answer quality alone. However, their latency behavior changes drastically with server load, as shown in Fig. \ref{fig:agentcompare}. Under light load, Self-Consistency achieves a lower average JCT than ReAct (20.3\,s vs.\ 26.9\,s), reducing average JCT by about 24.5\%. This is because its higher intra-agent parallelism can better exploit idle resources. Under heavy load, the trend reverses: ReAct achieves an average JCT of 29.5\,s, whereas Self-Consistency degrades to 82.5\,s. In this regime, the additional parallel branches in Self-Consistency compete more aggressively for limited inference capacity, creating substantial queueing delays and amplifying contention. This result shows that the "best" agent cannot be determined from its structure alone. A more complex agent may be latency-efficient when the server is lightly loaded, yet become significantly worse once resources are constrained. Such server-dependent behavior cannot be captured by static or server-agnostic cost models, and therefore makes server-aware adaptation necessary.

\subsection{Motivation for Probabilistic Agent Scheduling}
\label{sec:motivation_agentscheduling}
\textbf{Observation 3: Existing scheduling schemes induce severe HoL blocking in agentic workloads.}
{Agent runtimes are highly uncertain due to stochastic LLM generation, query-dependent tool use, and dynamic planning loops, so static schedulers cannot distinguish short-running agents from long-running ones before execution, causing long agents to block many shorter ones. We construct a long-tail mixed workload (60\% simple I/O agents, 30\% medium agents such as ReAct and Self-Consistency, and 10\% complex AFlow agents, all using Qwen3-4B) and compare Parrot \cite{parrot}, Ayo \cite{ayo}, Agentix \cite{agentix}, and an Oracle scheduler that prioritizes agents by their true service times, approximating ideal shortest-job-first scheduling.}

{As shown in Fig. \ref{fig:motivation3}, existing schedulers remain far from the Oracle: even the best one, Agentix, incurs nearly twice the average JCT. This gap stems not merely from workload skewness but from the inability to infer which agents will finish soon under execution uncertainty. Since exact service times are unavailable before execution, a scheduler should account for the distributional nature of agent runtimes rather than rely on deterministic size estimates alone.}

\subsection{Motivation for Topology-aware Model Scheduling}
\label{sec:motivation_modelscheudling}

\begin{figure}[t]

  \centering

    \includegraphics[width=\linewidth]{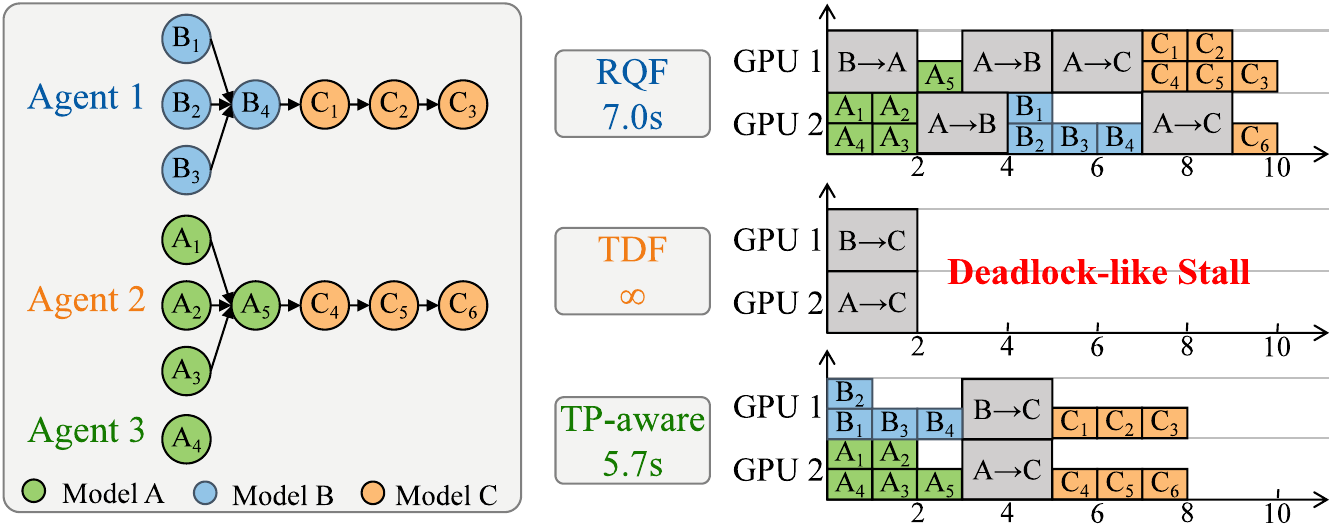}

            \caption{JCT comparison among RQF, TDF, and TP-aware.}

        \label{fig:motivation4}

\end{figure}

\textbf{Observation 4: Request-level model scheduling fails under graph-shaped agentic workloads.}
Stage dependencies determine when model demand becomes executable. Model scheduling must consider both the volume and release timing of pending requests. Fig.~\ref{fig:motivation4} presents a toy example with two GPUs, each serving up to two concurrent same-model requests. Agents 1 and 2 are complex AFlow agents combining Self-Consistency and ReAct; agent 3 is simple I/O. Each stage takes one time unit, and each model switch takes two. We consider two demand-driven policies: Ready-Queue-First (RQF) and Total-Demand-First (TDF). RQF counts only ready stages, repeatedly switching models as execution exposes new requests. This reactive behavior incurs loading overhead, increasing average JCT to 7\,s. TDF instead counts all remaining stages, including blocked ones. It loads C prematurely for its largest total demand, reducing the capacity available to unfinished A- and B-stages. The resulting deadlock-like stall prevents these upstream stages from unlocking most C-stages. In contrast, topology-aware (TP-aware) scheduling considers future demand and switching overhead together: it retains A/B capacity to unlock downstream C-stages, then switches as their demand approaches, achieving the lowest average JCT of 5.7\,s. Consequently, topology-awareness should be incorporated into model scheduling.

\begin{figure}[t]
  \centering
    \includegraphics[width=\linewidth]{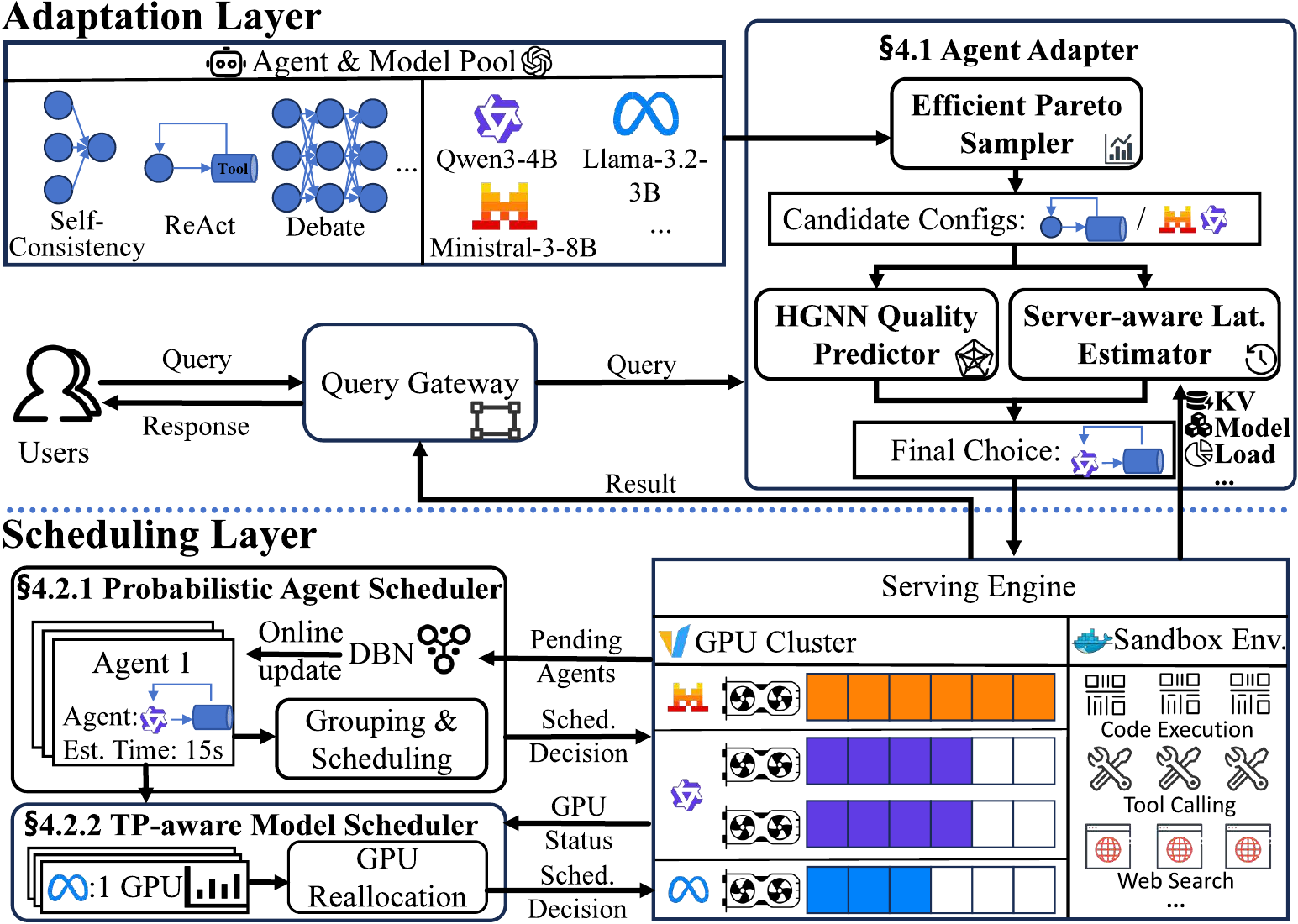}
            \caption{Overview of Presage.}
        \label{fig:framework}
\end{figure}

\section{Presage Framework}
\label{sec:framework}

We present an overview of Presage in Fig. \ref{fig:framework}. The goal of Presage is to minimize the average JCT while selecting configurations predicted to meet a per-query answer-quality target.
 It mainly consists of {an agent adaptation layer} (\S\ref{sec:agentadapter}) and {a predictive scheduling layer} (\S\ref{sec:scheduler}). When a query is submitted to Presage, the {adaptation layer} selects the most appropriate agent-model configuration to serve it based on query semantics and {server state}. Specifically, it uses an HGNN-based predictor (\S\ref{sec:hgnn_predictor}) to predict the answer quality given a specific agent and LLM(s), and a server-aware latency estimator (\S\ref{sec:server_aware_latency}) to estimate the E2E latency. The adapter then chooses the most appropriate agent-model configuration with an efficient Pareto sampling algorithm. Next, the agent is instantiated and sent to the {scheduling layer} for execution. {In the scheduling layer}, our probabilistic agent scheduler (\S\ref{sec:agentscheduler}) and topology-aware model scheduler (\S\ref{sec:modelscheduler}) make {predictive} scheduling decisions to minimize the average JCT.

\subsection{Server-aware Agent Adapter}
\label{sec:agentadapter}

\subsubsection{HGNN-based Quality Predictor}
\label{sec:hgnn_predictor}
{For online adaptation, Presage must estimate the answer quality of candidate configurations, which depends jointly on the user query, the agent workflow, and the models assigned to different stages. We therefore propose an HGNN-based quality predictor that explicitly models each candidate configuration as a heterogeneous graph and learns query-specific quality estimates $y \in [0,1]$; quality is measured by task-specific metrics such as pass@1 for code generation, F1 score for reading comprehension, and accuracy for math reasoning.}

\begin{figure}[t]
  \centering
    \includegraphics[width=\linewidth]{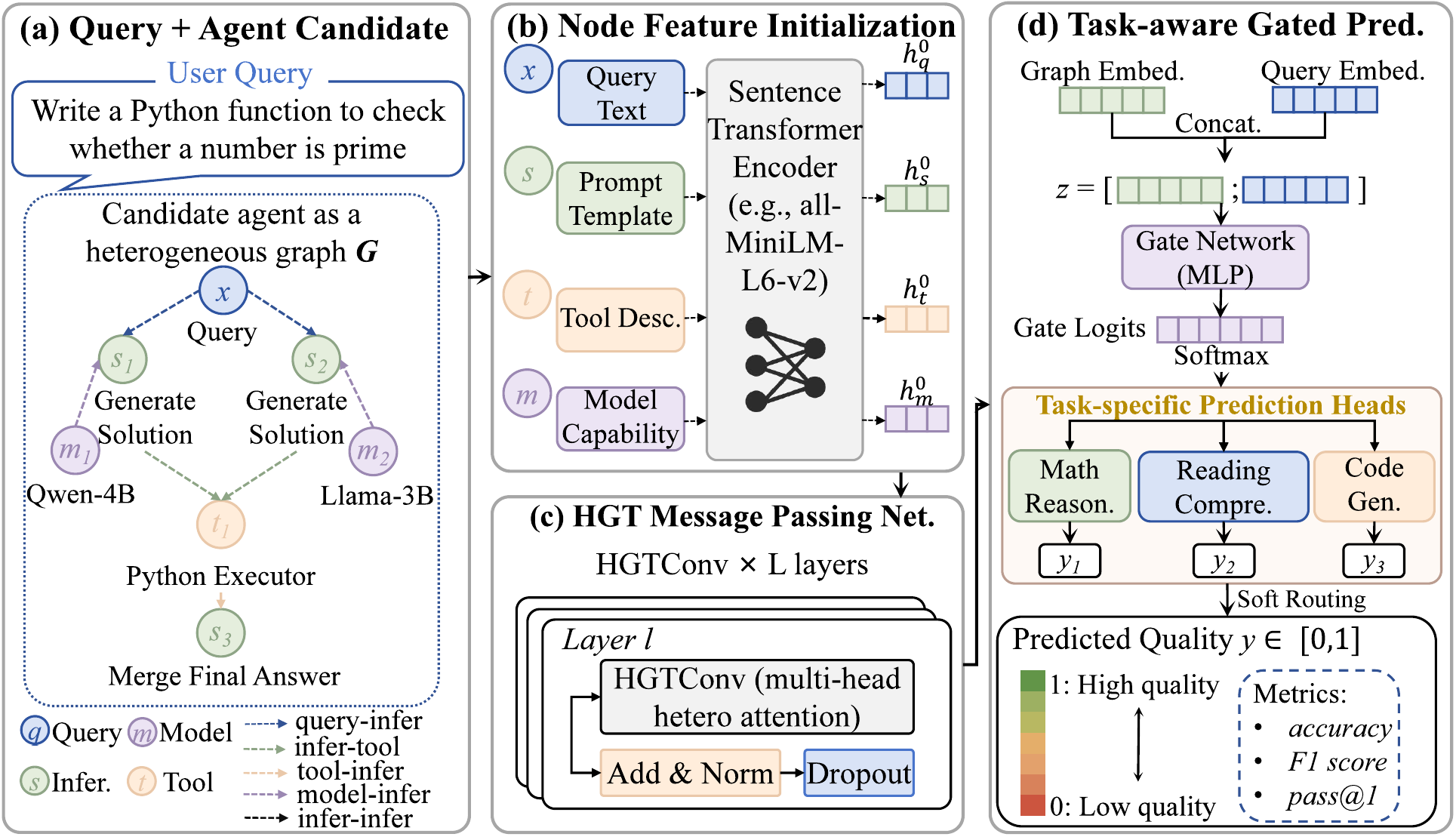}
            \caption{HGNN-based answer quality predictor.}
        \label{fig:gnn}
\end{figure}

{As shown in Fig. \ref{fig:gnn}, the graph contains query, inference, tool, and model nodes, with heterogeneous edges encoding execution order, model usage, and information flow. Node features are initialized by embedding textual descriptions or stage prompt templates with a pretrained model (all-MiniLM-L6-v2 \cite{Minilm}), where concise introductions of candidate LLMs and tools are generated by an LLM (GPT-5.6 in our paper).}

{Nodes are first projected into a shared latent space by type-specific encoders, followed by Heterogeneous Graph Transformer \cite{hgt} layers for type-aware message passing. We mean-pool the contextualized embeddings of all non-query nodes into an agent-level representation and concatenate it with the query-node embedding to capture both global agent context and query semantics.}

{To handle task heterogeneity, a lightweight gating network routes the concatenated agent-query representation to multiple task-specific prediction heads, and the final quality estimate $y \in [0, 1]$ is a soft mixture of all heads, which is used to guide agent adaptation.}

\subsubsection{Server-aware Latency Estimator}
\label{sec:server_aware_latency}

In addition to answer quality, a critical factor in serving LLM agents is the end-to-end (E2E) latency between query submission and final-response return. However, accurately estimating the E2E latency of an instantiated agent remains challenging: the realized latency is not determined solely by the agent graph but is also heavily modulated by the server state, including KV cache reuse, model residency, and workload intensity. Consequently, we propose a lightweight server-aware estimator that explicitly captures server-related factors.

{We summarize the server state at decision time as a snapshot $S=(S_{\mathrm{KV}}, S_{\mathrm{LLM}}, S_{\mathrm{W}})$, where $S_{\mathrm{KV}}$ captures KV cache reuse statistics under the agent's prompt template, $S_{\mathrm{LLM}}$ the GPU-resident model sets, and $S_{\mathrm{W}}$ per-model queueing statistics and effective serving capacity. Given an agent graph $G=(V,E)$, we estimate its E2E latency by simulating its stages over this snapshot, decomposing each stage's latency into queueing delay and service time, and further decomposing the service time of LLM stages into prefill and decode latency:}
\begin{equation}
l_v^{\mathrm{svc}} = l_v^{\mathrm{pre}} + l_v^{\mathrm{dec}}.
\end{equation}

{We first model KV cache reuse. For an LLM stage $v$, let $l_{v,\mathrm{full}}^{\mathrm{pre}}$ be the prefill latency without KV reuse and $s_v\in[0,1]$ the fraction of prompt tokens reusable from cache; the effective prefill latency follows a lightweight linear approximation:}
\begin{equation}
l_v^{\mathrm{pre}}
=
l_{v,\mathrm{full}}^{\mathrm{pre}} (1-s_v).
\label{eq:kv_prefill}
\end{equation}
{Since most serving engines do not expose exact reusable cached tokens, we estimate $s_v$ from cache-reuse evidence of recent requests with the same prompt template (e.g., the \textit{num\_cached\_tokens} reported by vLLM), incorporating KV effects without intrusive engine instrumentation.}

{We next model LLM residency, assuming homogeneous GPUs. For each model $m$, we estimate its ready time $T_m^{\mathrm{ready}}$, i.e., the earliest time $m$ can begin serving requests, which is $0$ if $m$ is already resident. Otherwise, a candidate GPU $g$ hosting model $m'$ must first drain its remaining inference workload and offload $m'$, so its idle time is:}
\begin{equation}
T_{g}^{\mathrm{idle}}
=
T_g^{\mathrm{busy}} + T_{m'}^{\mathrm{offload}},
\end{equation}
{where $T_g^{\mathrm{busy}}$ is the estimated remaining busy time of GPU $g$ and $T_{m'}^{\mathrm{offload}}$ the profiled offloading cost of $m'$. The ready time of $m$ then picks the GPU making $m$ available earliest:}
\begin{equation}
T_m^{\mathrm{ready}}
=
\min_{g \in \mathcal{G}}
\left(
T_{g}^{\mathrm{idle}} + T_{ m}^{\mathrm{load}}
\right),
\end{equation}
where $\mathcal{G}$ is the set of candidate GPUs and $T_{m}^{\mathrm{load}}$ denotes the profiled cost of loading model $m$ onto a GPU.

We further model the effect of workload intensity. For a resident model $m$, let $b_m$ denote the average number of idle batch slots observed over a recent time window. We use $b_m$ as a lightweight estimate of residual capacity for LLM inference. If $b_m \geq 1$, the model still has spare capacity and we set per-stage queueing delay to $0$; otherwise, we use the running average queueing delay $\bar{q}_m$ from recently served requests. Thus, the per-stage queueing delay is estimated as $\bar{q}_m \cdot \mathbf{1}[b_m < 1]$, and the residual capacity as $\max(b_m,1)$ to avoid zero capacity.
For a non-resident model $m$, we estimate its residual capacity as $\max(C_{m}-r_{m},1)$, where $C_{m}$ denotes its batch capacity and $r_{m}$ denotes the expected number of concurrently running requests. Following Little’s Law \cite{little}, we estimate $r_{m} = \lambda_m p_m s_m$, where $\lambda_m$ is the arrival rate of agents requiring ${m}$, $p_m$ is the average parallelism of current agents on ${m}$, and $s_m$ is the empirical service time of ${m}$'s stages. It serves as a lightweight approximation to the topology-induced stage concurrency on model $m$ of each agent. Regarding the queueing delay, we estimate it only for the entry stage through a simple list-scheduling procedure, which computes the time needed to drain the backlog of currently ready stages requiring $m$.

{Given the above estimates, we run a lightweight event-driven simulation over the agent graph in topological order, where each model carries its estimated residual capacity and ready time: a stage starts once all its predecessors complete, its required model is ready, and its queueing delay has elapsed, and the finish time of the final sink stage gives the estimated E2E latency.}

\subsubsection{Pareto-sampling-based Agent Adaptation}
\label{sec:pareto_sampling}
With the quality predictor and the latency estimator, we next design an efficient component-level adaptation algorithm. Instead of assigning models at the whole-agent or individual-stage level as in existing works \cite{masrouter,routereason}, Presage identifies and treats each reusable simple agent inside a complex agent as a base component for model assignment. This design is based on the observation that complex agents are often compositions of a few reusable simple agents, such as ReAct and Self-Consistency in AFlow \cite{aflow} and MaAS \cite{maas}. It provides a middle ground between agent-level assignment, which is too coarse to capture functional heterogeneity, and per-stage assignment, whose combinatorial search space is impractical for online serving. Based on this design, Presage first performs a coarse-grained scan with uniform model assignments to capture the agent-level quality--latency trade-off, and then refines the search through Pareto-guided component-level sampling around the target quality. This leads to the following two-step adaptation algorithm.

a) Step 1: coarse-grained enumeration. We first enumerate all agents together with all candidate models under a uniform assignment. That is, for each agent, we evaluate the configuration in which all stages use the same model. This step serves as a cheap global scan over the candidate space and provides a common reference point for all agents under the current query and server state. For simple agents, this step already exhausts their search space, since each simple agent contains only one component. For complex agents, although uniform assignments cannot capture all useful mixed-model configurations, they provide a coarse approximation of the agent-level trade-off, and such information will be used to further generate promising candidates in the next step.

b) Step 2: efficient Pareto sampling. For each complex agent, we construct its agent-level Pareto frontier from the uniform assignments evaluated in step 1. Each point on the frontier corresponds to a unique model. We then choose two anchor points on this frontier to induce a compact search region for each component. The first is a latency-oriented anchor, defined as the Pareto point with quality immediately below $q_{\mathrm{tar}}$. It represents the cheapest near-feasible configuration. The second is a quality-oriented anchor whose predicted quality is higher than but closest to $q_{\mathrm{tar}}$. Together, these two anchors define a compact search range. If either anchor does not exist, we skip the current agent and move to the next one.

Next, we use these two anchors to induce a model subset for each component in the complex agent. Specifically, for each reusable component, we map the two anchor models to their corresponding simple-agent profiles obtained in step 1, and use the resulting component-level quality and latency values to define a rectangular search region in the quality-latency plane. We focus on models that lie inside and remain Pareto-optimal within the region, and use these models as the candidate pool for the current component. In this way, we induce a compact and target-adaptive search area for each component, reducing the search space while preserving the most relevant trade-off candidates.

{With the reduced search space, we sample at most $K$ candidates and evaluate them in parallel using the quality predictor and latency estimator. Among candidates satisfying the quality target, we select the one with the lowest estimated latency; if none qualifies, we fall back to the candidate with the highest predicted quality. Algorithm~\ref{alg:agent_adaptation} in Appendix~\ref{app:algorithms} gives the full procedure.
}

\subsection{Predictive Scheduling Engine}
\label{sec:scheduler}

\subsubsection{Probabilistic Agent Scheduler}
\label{sec:agentscheduler}
Mitigating head-of-line blocking and increasing system responsiveness  requires the scheduler to identify and prioritize short-running agents.
However, the stochastic nature of LLM generation and the dynamic execution paths of agents make it difficult to obtain precise resource demand and execution topology. To address this challenge, we propose a probabilistic agent scheduler. It is based on the observation that LLM agents typically operate iteratively and exhibit strong temporal and spatial correlations across stages. Specifically, stages within the same iteration or parallel branch may share similar inputs, prompt structures, or intermediate context, while stages across iterations are often coupled through accumulated context and execution history, leading to correlated prompt/completion lengths and execution durations. We present the supporting measurements of stage-level spatial and temporal correlations for commonly-used agents in Appendix~\ref{app:correlation}.

Inspired by the above observation, we leverage the Dynamic Bayesian Network (DBN) \cite{dbn} to capture such temporal and spatial correlations and propose a probabilistic agent scheduler. A DBN is a probabilistic graphical model that represents and reasons about sequences of variables evolving over time. Given the observed runtime information of early stages, the DBN updates its posterior over downstream stage variables, enabling online prediction of future execution behavior and the remaining duration of the agent.
{Nodes represent token variables for LLM stages (prefill/decode tokens) and duration variables for regular stages.}

At runtime, we incorporate observed token counts and stage durations
as evidence to update the DBN posterior. For LLM stages, posterior
token-length predictions are mapped to service times using the
profiled token-to-time relationship, while non-LLM stages are
represented directly by duration variables. We use posterior means
to obtain stage-level duration estimates and combine them according
to the dependencies of the remaining workflow to estimate the
agent's remaining duration $\hat{R}_a$. As execution evidence
accumulates, these estimates are updated, allowing scheduling
priorities to reflect the progress of individual agents. We partition agents into multiple groups according to model residency status. Specifically, each group corresponds to the number of required models for an agent that is currently non-resident. During scheduling, we consider the groups in order, from agents requiring the fewest additional model loads to those requiring the most. Within each group, we rank agents by the DBN-estimated remaining duration, and for each agent, we further prioritize the stage lying on its residual critical path (CP), sequentially admitting the stage into the LLM's inference batch or sandbox execution. To avoid starvation, we add a simple aging mechanism by replacing the ranking key $\hat{R}_a$ with $\hat{R}_a-\alpha W_a$, where $W_a$ denotes the waiting time, and $\alpha$ is a small aging coefficient that can be flexibly adjusted to trade off responsiveness and fairness. The full procedure is given in Algorithm~\ref{alg:prob_job_scheduler} in Appendix~\ref{app:algorithms}.

\subsubsection{Topology-aware Model Scheduler}
\label{sec:modelscheduler}
{Since the graph-structured execution of LLM agents hides future model demand from the current ready queue, we design a topology-aware model scheduler that combines the agent topology with the DBN-based runtime estimates and reallocates GPUs from over-provisioned models to under-provisioned models. For each active agent, we obtain each stage's expected execution duration conditioned on the current evidence. For a stage $v$ with predecessor set $\mathrm{pred}(v)$, the timestamp it becomes ready for execution is:}
\begin{equation}
R_v=\max_{u\in \mathrm{pred}(v)} F_u,
\end{equation}
where $F_u$ denotes the completion time of predecessor $u$. Discretizing the horizon $H$ into buckets of width $\Delta$, the estimated concurrent demand for model $m$ in bucket $k$ is:
\begin{equation}
\label{eq:model_demand_curve}
D_m[k]
= \sum_{v:\,\mathrm{model}(v)=m}
\kappa_v^k,
\end{equation}
where $\kappa_v^k$ is a binary indicator specifying whether stage $v$ is ready for execution in bucket $k$.
{This captures topology-induced future fan-out that is invisible from the current ready set, with $\Delta$ set dynamically from a running average of recent inference service times.}

{Given the demand curve, models are classified as over-provisioned or under-provisioned depending on whether their assigned GPUs can sustain the peak parallel demand over the lookahead horizon. Under-provisioned models whose peak unmet demand emerges earlier are prioritized as likely near-term bottlenecks, while over-provisioned models with the largest surplus capacity donate GPUs first. To select a specific GPU for switching, we rank GPUs by the estimated remaining service time of their current batches in ascending order, so that reallocation can be enacted as early as possible. Algorithm~\ref{alg:topo_model_scheduler} in Appendix~\ref{app:algorithms} gives the full procedure.
}

\section{Evaluation}
\label{sec:evaluation}

\begin{figure*}[t]
    \centering
     \includegraphics[width=1.0\textwidth]{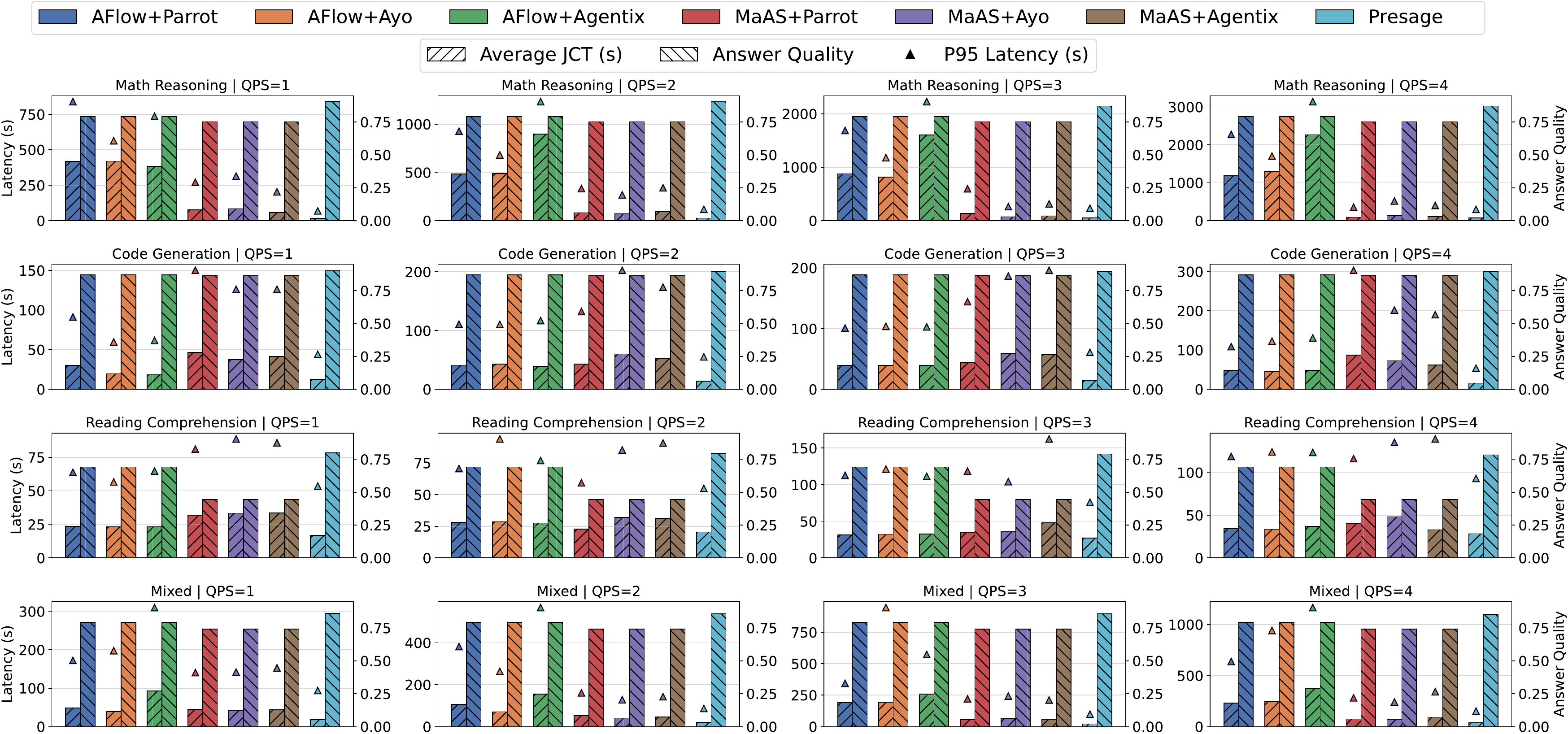}
    \caption{Average JCT (left bar), answer quality (right bar), and P95 latency (triangle markers) on different query arrival rates and agent workloads.}
    
    \label{fig:overall}
\end{figure*}

\noindent\textbf{Implementation.}
{We implement Presage in Python with approximately 5K lines of code, serving LLMs with vLLM \cite{vllm} on a server with four 48GB H800 MIG instances obtained from two H800-96GB GPUs \cite{mig}. The HGNN-based predictor uses a 256-dimensional hidden state, 4 attention heads, and 2 layers, trained with soft-label binary cross-entropy loss and AdamW at a learning rate of $2\times10^{-3}$. We maintain a DBN per agent for the joint prior over stage-level variables, and select the estimator window size, sampling budget $K$, and horizon length $H$ via constrained grid search on a held-out validation trace.}

\noindent\textbf{Models and agents.} For LLMs, we use the same six models mentioned in \S \ref{sec:motivation}. For agents, we use I/O, ReAct, Self-Consistency, Debate, and complex agents constructed by AFlow \cite{aflow}, totaling 9 different agents.

\noindent \textbf{Workload.} {We construct workloads by scaling the timestamps from the Azure inference trace~\cite{llmtrace} to different arrival rates and replaying them as request arrivals, using queries from the following benchmarks: MBPP \cite{mbpp} and HumanEval \cite{humaneval} for code generation; MATH \cite{mathdataset} and GSM8K \cite{gsm8k} for math reasoning; HotpotQA \cite{hotpotqa} and DROP \cite{drop} for reading comprehension; and a mixed workload evenly mixing queries from the six datasets.}

\noindent\textbf{Baselines.} {Since, to the best of our knowledge, there are no end-to-end adaptive agent serving frameworks, we build six E2E baselines by combining SOTA adaptation methods with agent serving systems. For the adaptation method, we choose AFlow \cite{aflow}, which generates task-level agents via offline Monte Carlo Tree Search and adapts them to queries by task type, and MaAS \cite{maas}, which samples query-dependent agents from a well-trained supernet, with per-token prices from OpenRouter \cite{openrouter}. For the serving system, we choose Parrot \cite{parrot}, which abstracts semantic variables from LLM agents and co-locates prefix-sharing agents on the same inference engine; Ayo \cite{ayo}, which represents agents as primitive-level dataflow graphs and orchestrates their execution with topology-aware batching; and Agentix \cite{agentix}, which depicts each agent as a general program and mitigates HoL blocking with a multi-level program-based scheduler. We use 80\% of the data from each dataset for training all trainable modules in Presage and other baselines, and the remaining 20\% for evaluation.}

\noindent\textbf{Evaluation Metrics.}
{We evaluate answer quality using task-specific metrics: F1 score for reading comprehension, pass@1 for coding, and accuracy for math reasoning, averaging offline evaluations over five random seeds per agent-model configuration to account for generation stochasticity. For serving latency, we report the average and P95 JCT.}

\begin{figure}[t]
  \centering
    \includegraphics[width=\linewidth]{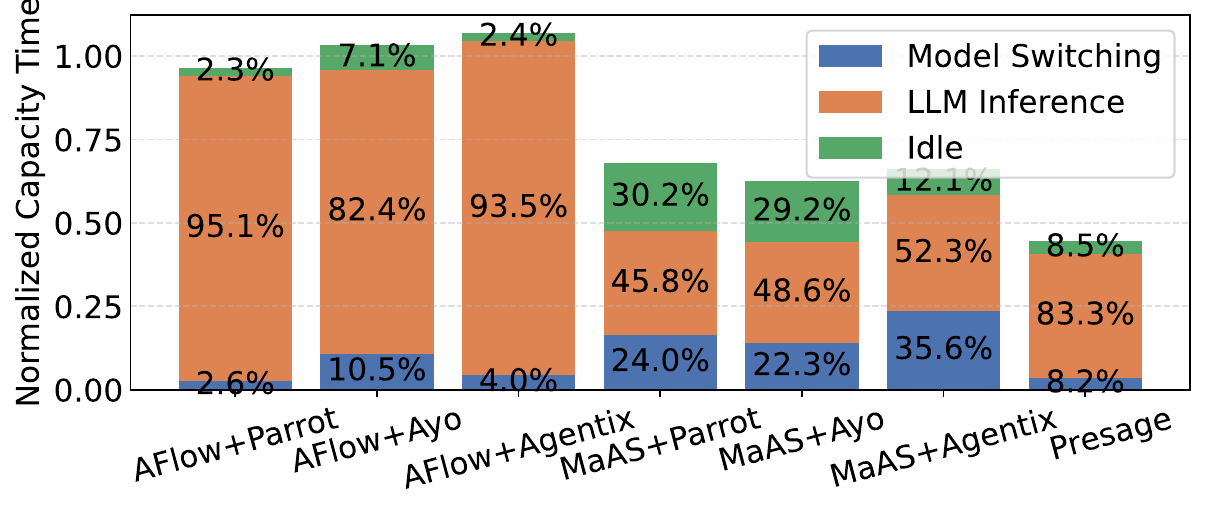}
            \caption{Runtime breakdown of baselines and Presage.}
        \label{fig:breakdown}
\end{figure}

\subsection{Overall Results}

We first evaluate the average answer quality and JCT across different workload types and query arrival rates. To achieve competitive answer quality and account for the performance heterogeneity across benchmarks, we set the default quality target for each benchmark to the lowest answer quality among the top 20\% of all evaluated agent-model configurations. As shown in Fig. \ref{fig:overall}, Presage consistently achieves the lowest average JCT across workloads and the highest answer quality. Compared with other baselines, Presage reduces average JCT by 32.3--81.9\%, 19.2--97.1\%, 48.1--91.7\%, and 10.4--49.9\% on math reasoning, code generation, reading comprehension, and mixed workloads, respectively. Regarding the tail performance, Presage reduces P95 latency by 11.7--91.8\%, 26.1--82.2\%, 32.7--89.5\%, and 5.5--53.2\%, respectively. Importantly, these latency reductions do not come at the cost of answer quality. Instead, Presage improves answer quality by 3.1--4.2\%, 9.9--20.6\%, 7.6--16.2\%, and 13.3--80.3\% on the four workloads, respectively.

{To further understand the source of improvement, we present the runtime breakdown on the mixed workload with queries per second (QPS) = 3 in Fig. \ref{fig:breakdown}. AFlow-based baselines select complex and heavyweight agents regardless of server load, overloading the LLM engines and causing long queueing delays. MaAS reduces system pressure by considering token usage, but token efficiency alone is not a reliable proxy for serving efficiency: it blindly chooses among models regardless of residency, causing frequent model switching, limited latency improvement, and substantial quality drops.}

{The serving engine further affects E2E performance: Parrot suffers from severe HoL blocking under heterogeneous agent workloads; Ayo exploits agent topology but ignores the duration distribution of dynamic agent stages; Agentix's MLFQ-style mechanism is reactive and insufficient for uncertain agent executions. In contrast, Presage avoids unnecessarily heavy agents, reduces model switching overhead, and prioritizes short-running agents, achieving substantially lower JCT while improving answer quality.}

{We also evaluate Presage's HGNN-based quality predictor against several commonly used alternatives, including an off-the-shelf homogeneous GNN, a deep feedforward network (FFN), support vector regression (SVR), and decision trees (DT). As shown in Table \ref{tab:predictor}, our predictor consistently achieves the lowest prediction error across all tasks, confirming the benefit of modeling agents as heterogeneous graphs with task-aware prediction heads and enabling more reliable adaptation decisions.}

\begin{table}[t]

\centering
\small
\setlength{\tabcolsep}{2.6pt}
\renewcommand{\arraystretch}{1.08}
\begin{threeparttable}
\caption{Answer quality prediction error ($\times 10^{-2}$) of different methods.}
\label{tab:predictor}
\begin{tabular}{l*{8}{c}}
\toprule
\multirow{2}{*}{Method}
& \multicolumn{2}{c}{Math}
& \multicolumn{2}{c}{Reading}
& \multicolumn{2}{c}{Coding}
& \multicolumn{2}{c}{All} \\
\cmidrule(lr){2-3}
\cmidrule(lr){4-5}
\cmidrule(lr){6-7}
\cmidrule(lr){8-9}
& MSE & MAE & MSE & MAE & MSE & MAE & MSE & MAE \\
\midrule
GNN & 1.68 & 7.45 & 0.81 & 5.63 & 2.11 & 8.45 & 1.64 & 7.18 \\
FFN & 4.49 & 13.25 & 3.47 & 12.44 & 4.96 & 15.00 & 4.07 & 13.28 \\
SVR & 19.81 & 37.68 & 9.78 & 22.87 & 18.54 & 32.29 & 12.80 & 25.91 \\
DT & 13.51 & 23.04 & 6.13 & 19.21 & 14.24 & 27.38 & 12.05 & 24.03 \\
\textbf{Ours} & \textbf{0.85} & \textbf{4.86} & \textbf{0.45} & \textbf{3.89} & \textbf{1.02} & \textbf{5.43} & \textbf{0.95} &  \textbf{5.11} \\
\bottomrule
\end{tabular}
\end{threeparttable}
\end{table}

\subsection{Ablation Study}
We ablate the key components of Presage to understand their contributions. Unless otherwise specified, all results are obtained on the mixed workload under QPS=3.

\begin{figure}[t]
    \centering
    \subfigure[Light load]{
        \includegraphics[width=0.22\textwidth]{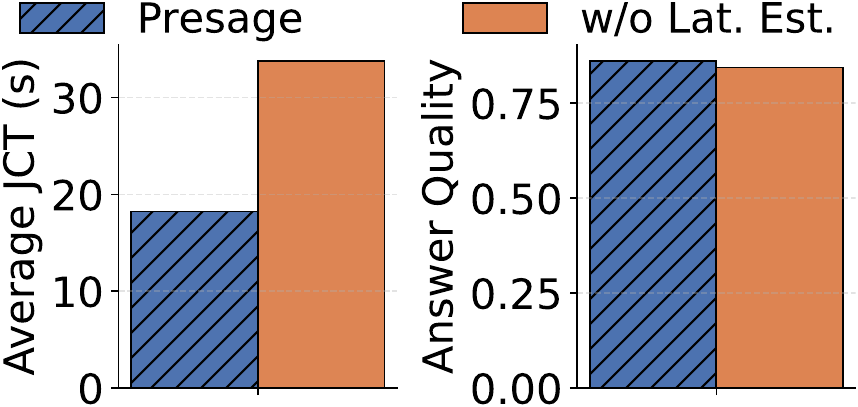}
        \label{fig:abl_latencymodel_lightload}
    }
    \subfigure[Heavy load]{
        \includegraphics[width=0.22\textwidth]{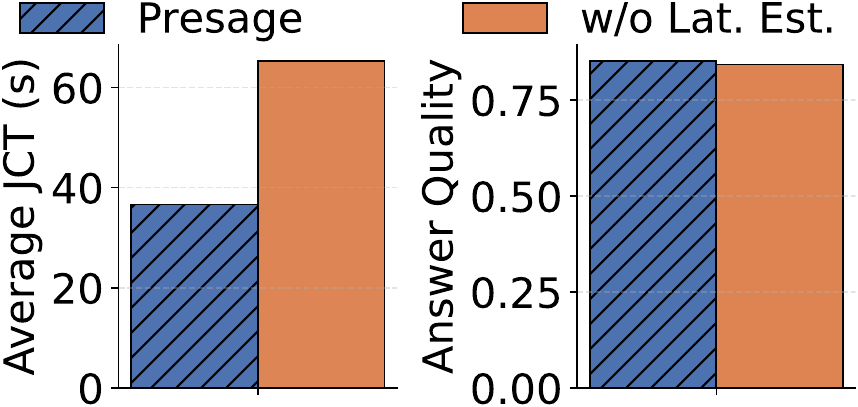}
        \label{fig:abl_latencymodel_heavyload}
    }
    \caption{Ablation study on server-aware latency estimator.}
    \label{fig:abl_latencymodel}
\end{figure}

\noindent\textbf{Effectiveness of Server-aware Latency Estimator.}
{We remove the server-aware latency estimator and construct a baseline named \textit{w/o Lat. Est.}, which estimates latency only from offline LLM profiles and agent structure. As shown in Fig. \ref{fig:abl_latencymodel}, Presage reduces average JCT by 46.1\% under light load (QPS=2) and 43.9\% under heavy load (QPS=4), while improving answer quality by 2.1\% and 1.2\%, respectively, showing that runtime states such as model residency, queueing pressure, model switching, and KV-cache reuse are essential for selecting agents that are both accurate and efficiently executable.}

\begin{figure}[t]
    \centering
    \subfigure[Average JCT and answer quality.]{
        \includegraphics[width=0.22\textwidth]{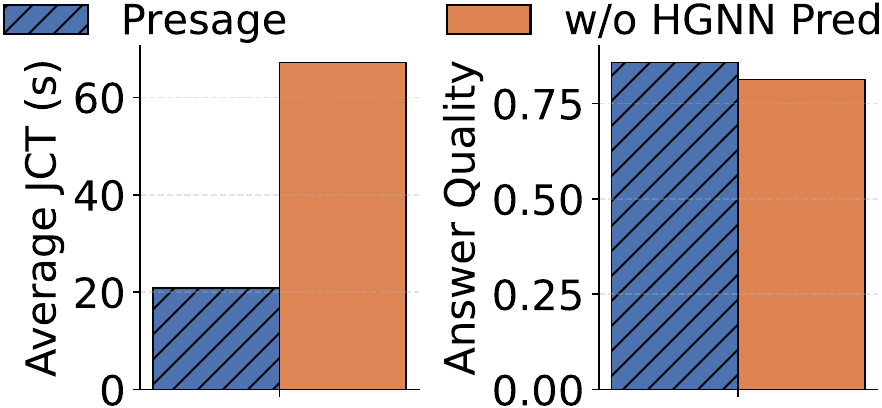}
        \label{fig:abl_hgnn_bar}
    }\hfill
    \subfigure[Quality gap distribution of target-violated queries.]{
        \includegraphics[width=0.22\textwidth]{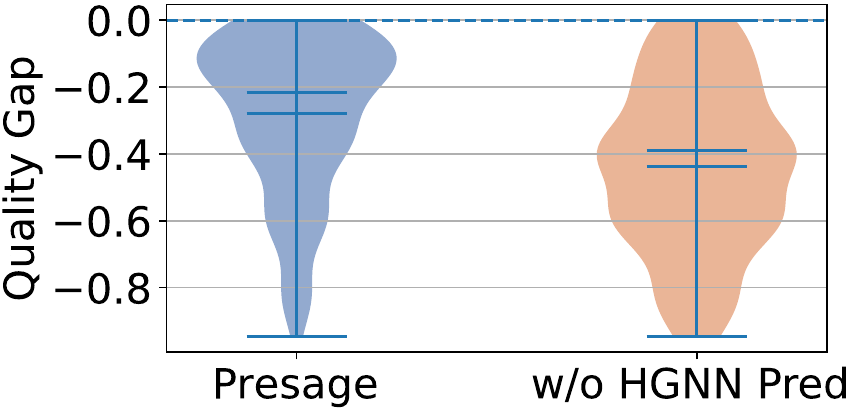}
        \label{fig:abl_hgnn_violin}
    }
    \caption{Ablation study on HGNN-based quality predictor.}
    \label{fig:abl_hgnn}
\end{figure}

\noindent\textbf{Effectiveness of HGNN-based Quality Predictor.}
{We replace the HGNN-based quality predictor with a coarse benchmark-level mean quality estimate, denoted as \textit{w/o HGNN Pred}. As shown in Fig. \ref{fig:abl_hgnn_bar}, our predictor reduces average JCT by nearly 69.0\% and improves answer quality by 5.5\%, since it captures query-specific differences while coarse estimates may over-provision easy queries or under-provision difficult ones. Fig. \ref{fig:abl_hgnn_violin} further shows a smaller and more compact quality-gap (i.e., achieved quality - target quality) distribution for target-violated queries, confirming more reliable quality estimates.}

\begin{figure}[t]
    \centering
    \subfigure[Agent scheduler.]{
        \includegraphics[width=0.22\textwidth]{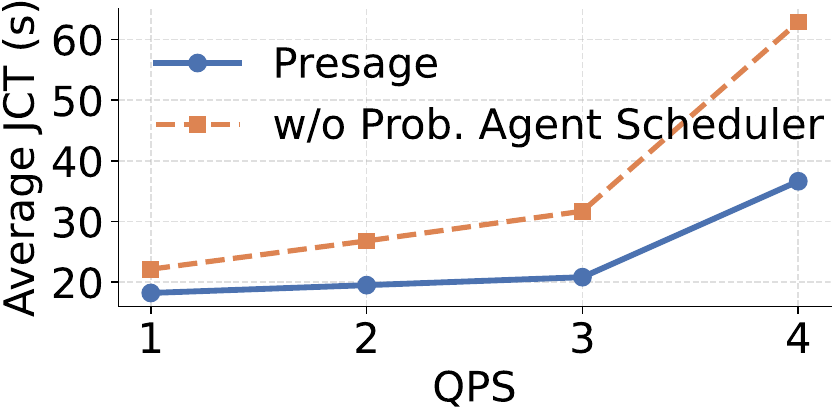}
        \label{fig:abl_jobscheduler}
    }
    \subfigure[Model scheduler.]{
        \includegraphics[width=0.22\textwidth]{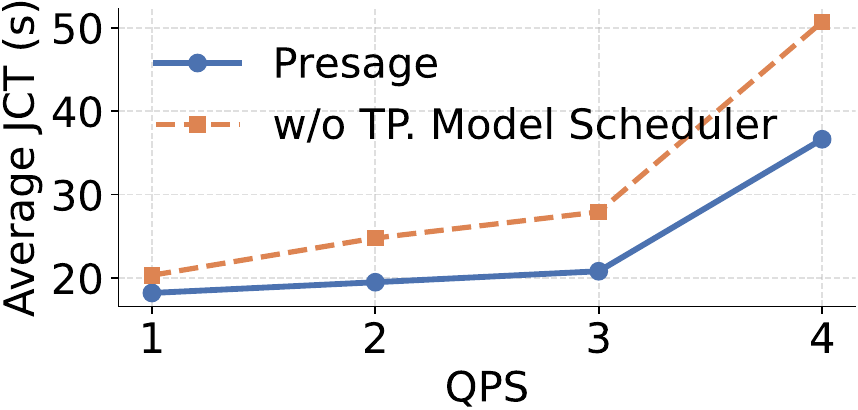}
        \label{fig:abl_modelscheduler}
    }
    \caption{Ablation study on agent and model schedulers.}
    \label{fig:abl_jobmodelscheduler}
\end{figure}

\begingroup\clubpenalty=10000
\noindent\textbf{Effectiveness of Probabilistic Agent Scheduler.}
{Next, we replace the DBN-based probabilistic agent scheduler with a naive FCFS scheduler, denoted as \textit{w/o Prob. Agent Scheduler}. As shown in Fig. \ref{fig:abl_jobscheduler}, Presage reduces average JCT by 17.7\%, 27.2\%, 34.3\%, and 41.8\% from QPS=1 to QPS=4. The gain increases under heavier load because FCFS suffers from HoL blocking, whereas Presage updates remaining-time estimates with stage-level evidence and prioritizes short-running agents and critical-path stages.}
\par\endgroup

\noindent\textbf{Effectiveness of Topology-aware Model Scheduler.}
{Finally, we replace the topology-aware model scheduler with a greedy scheduler that switches models based only on current ready requests, denoted as \textit{w/o TP. Model Scheduler}. As shown in Fig. \ref{fig:abl_modelscheduler}, Presage reduces average JCT by 10.4\%, 21.3\%, 25.5\%, and 27.7\% under QPS=1, 2, 3, and 4, respectively, showing that model scheduling must consider topology-dependent future demand rather than only the current ready queue, which reduces mistimed model swaps especially under high load.}

\begin{figure}[t]
  \centering
    \includegraphics[width=0.94\linewidth]{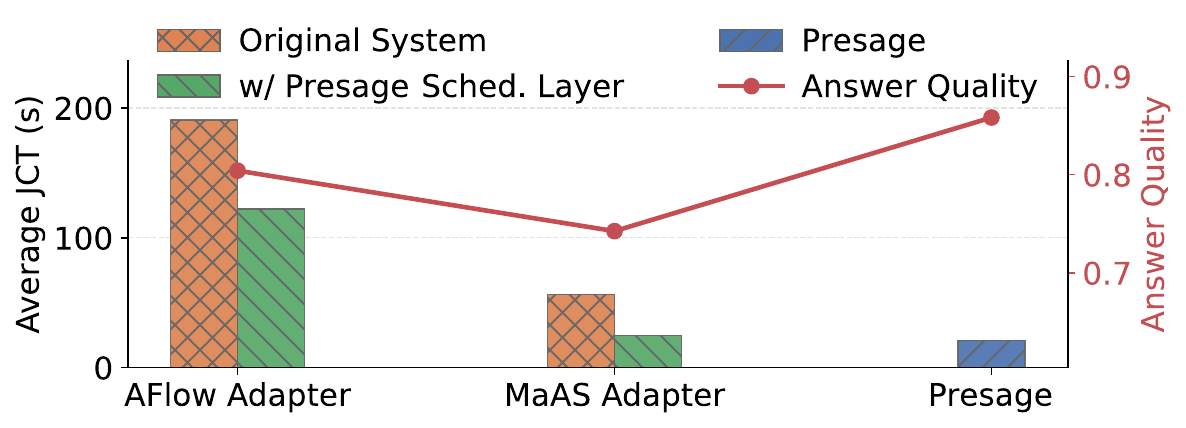}
            \caption{Effectiveness of Presage’s {scheduling layer} with different {adaptation methods}.}
        \label{fig:ablation_backend}
\end{figure}

{We further evaluate whether Presage's scheduling layer benefits existing adaptation methods. As shown in Fig. \ref{fig:ablation_backend}, replacing the original scheduling layer with Presage's reduces JCT by 35.9\% for AFlow and 56.1\% for MaAS, showing that the scheduling layer also improves externally generated agents. Overall, these results demonstrate the effectiveness of Presage's main components.}

\begin{table}[htbp]

\caption{Module overhead (ms) per invocation.}
\centering
\small
\begin{tabular}{lcccc}
\toprule
Overhead & QPS=1 & QPS=2 & QPS=3 & QPS=4 \\
\midrule
Agent Scheduler & 0.009 & 0.014 & 0.022 & 0.041 \\
Model Scheduler & 0.047 & 0.075 & 0.118 & 0.233 \\
Agent Adapter & 9.824 & 11.829 & 12.379 & 14.504 \\
DBN Update & 0.098 & 0.147 & 0.210 & 0.185 \\
\bottomrule
\end{tabular}

\label{tab:overheadanalysis}
\end{table}

\subsection{Sensitivity Analysis}
\label{sec:sensitivity}

\begin{figure}[t]

    \centering
    \subfigure[Quality targets.]{
        \includegraphics[width=0.46\textwidth]{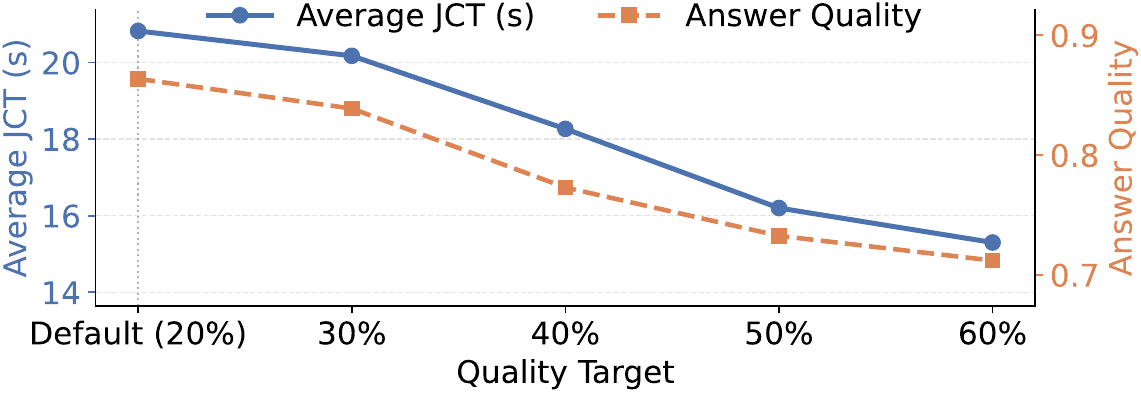}
        \label{fig:sens_target}
    }\hfill
    \subfigure[Number of model candidates.]{
        \includegraphics[width=0.22\textwidth]{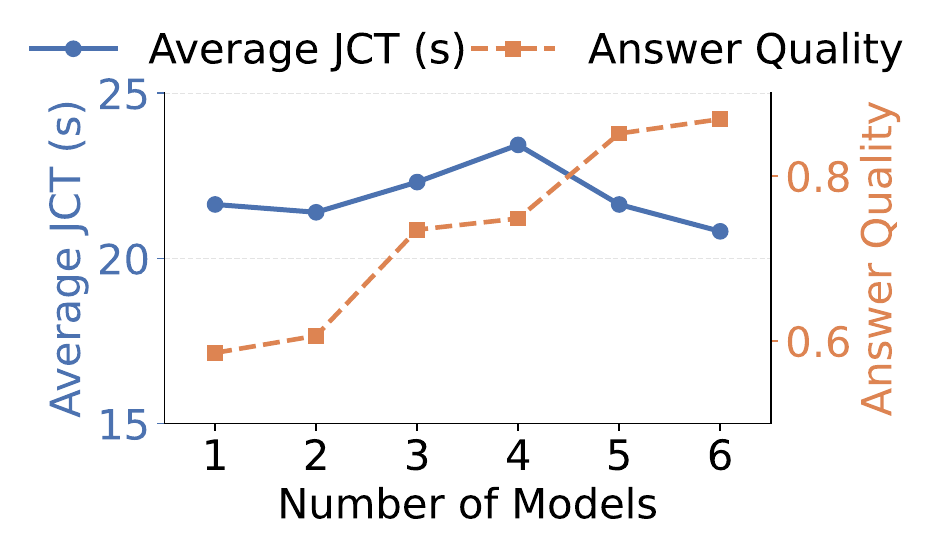}
        \label{fig:sens_model}
    }\hfill
    \subfigure[Number of agent candidates.]{
        \includegraphics[width=0.22\textwidth]{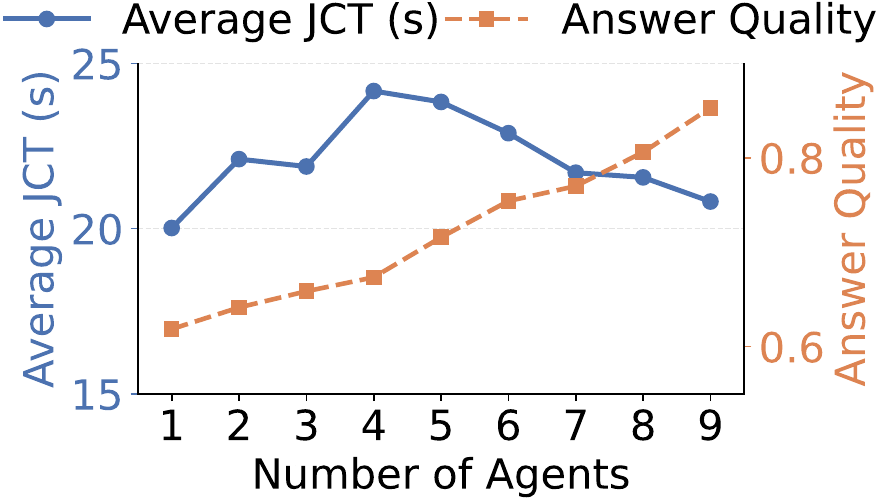}
        \label{fig:sens_agent}
    }
    \caption{Sensitivity analysis.}
    \label{fig:sensitivityanalysis}
\end{figure}

\textbf{Answer Quality Targets.}
{We evaluate Presage under answer quality targets from strict top-20\% to relaxed top-60\% on the mixed workload with QPS=3. In Fig. \ref{fig:sens_target}, Presage shows a clear and controllable quality--latency trade-off: the strict top-20\% target yields the highest answer quality of 86.3\% with the largest average JCT of 21.1~s due to stronger models or more complex agents, while relaxing the target to 60\% selects more lightweight configurations, reducing average JCT to 15.3~s with answer quality of 71.2\%. }

\noindent\textbf{Number of Agents and Models.}
{We evaluate the sensitivity of Presage to the number of models and agents, added one by one according to their release dates to mimic continuous deployment in production. As shown in Fig. \ref{fig:sens_model} and Fig. \ref{fig:sens_agent}, Presage consistently achieves higher answer quality as the pool grows. Although the average JCT may occasionally increase modestly when newly added models or agents are more accurate but heavier, each selected point remains Pareto-efficient: a higher latency is always accompanied by improved answer quality. }

\subsection{Overhead Analysis}
{We analyze the runtime overhead of Presage on the mixed workload under different QPS, as shown in Table \ref{tab:overheadanalysis}. Across QPS=1 to QPS=4, the agent scheduler takes only 0.009--0.041~ms and the model scheduler 0.047--0.233~ms, since both operate on compact runtime states; the DBN update is also lightweight at 0.098--0.210~ms. The largest overhead comes from the agent adapter (9.824--14.504~ms), which evaluates candidate assignments with both predictors, yet remains negligible compared with the tens or hundreds of seconds of agent service time, demonstrating that Presage is practical for real-time serving.}

\section{Related Work}
\label{sec:relatedwork}
\noindent\textbf{Agent Optimization.}
Many efforts optimize agent performance or cost through agent orchestration or routing \cite{masrouter,dyflow,gdesigner}. MasRouter \cite{masrouter} jointly routes LLMs and agents to minimize cost while maximizing answer quality. DyFlow \cite{dyflow} adjusts reasoning procedures based on task requirements and intermediate feedback. G-Designer \cite{gdesigner} designs task-adaptive communication topologies with a GNN. However, these works optimize only answer quality or token cost while remaining agnostic to server state, causing significant system inefficiency. Presage instead incorporates runtime server state into agent-model selection to reduce end-to-end serving latency.

\noindent\textbf{LLM Serving.}
Efficient LLM serving has been widely studied due to the proliferation of LLMs \cite{orca,flexgen,distserve}. 
Orca \cite{orca} proposes iteration-level scheduling and performs selective batching for LLM inference requests. 
FlexGen \cite{flexgen} enables large language models to run on a single commodity GPU by coordinating computation and data movement across GPU, CPU, and disk memory. 
DistServe \cite{distserve} disaggregates the prefill and decoding computation of LLM inference to different GPUs.
While these works improve LLM inference, they mainly target request-level optimization. Presage is orthogonal and complementary, focusing on agent adaptation and scheduling.

\begin{samepage}
\section{Conclusion}
\label{sec:conclusion}
In this paper, we present Presage, an E2E agent serving system with server-aware adaptation and predictive scheduling. By jointly considering answer quality, runtime server state, and agent execution topology, Presage leverages an HGNN-based quality predictor and a server-aware latency estimator to adaptively choose suitable agent-model configurations for each query. It further improves agent execution efficiency with a probabilistic agent scheduler and a topology-aware model scheduler. Experimental results show that Presage substantially reduces the E2E latency while maintaining or improving answer quality across diverse agentic workloads.
\par
\end{samepage}
\label{endofbody}
\clearpage
\bibliographystyle{mlsys2025}

\bibliography{references}

\clearpage
\setcounter{figure}{0}
\setcounter{algorithm}{0}
\appendix
\renewcommand{\thefigure}{A\arabic{figure}}
\renewcommand{\thealgorithm}{A\arabic{algorithm}}
\makeatletter
\renewcommand{\p@subfigure}{\thefigure}
\makeatother
\section{Stage-level Spatial and Temporal Correlations}
\label{app:correlation}
\begin{figure}[H]

    \centering

        \subfigure[ReAct agent.]{\includegraphics[width=0.48\columnwidth]{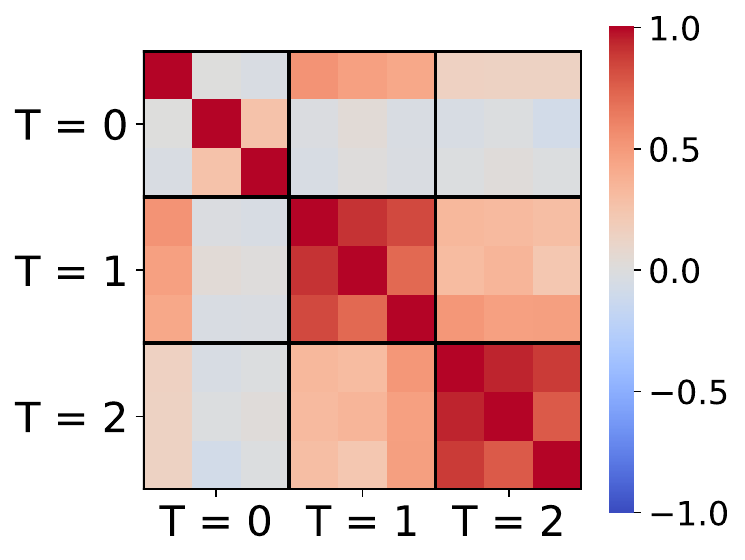}\label{fig:react correlation}}
    \hfill
    \subfigure[Self-Consistency agent.]{\includegraphics[width=0.48\columnwidth]{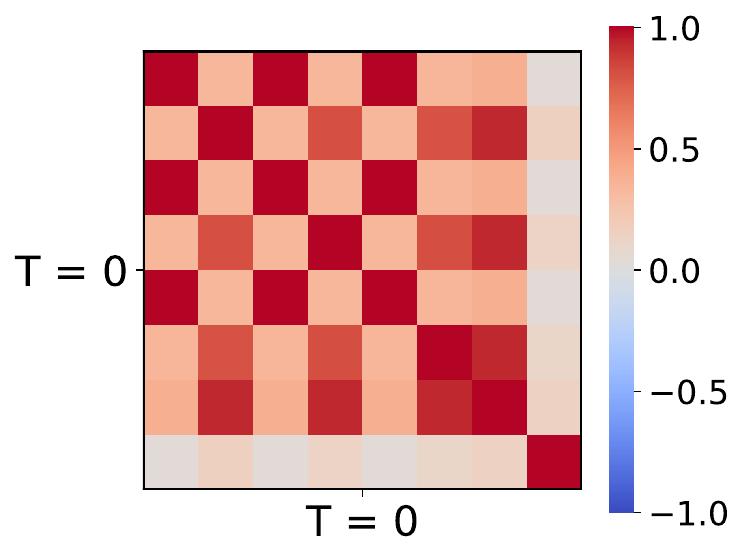}\label{fig:sc correlation}}
        \subfigure[Debate agent.]{\includegraphics[width=0.48\columnwidth]{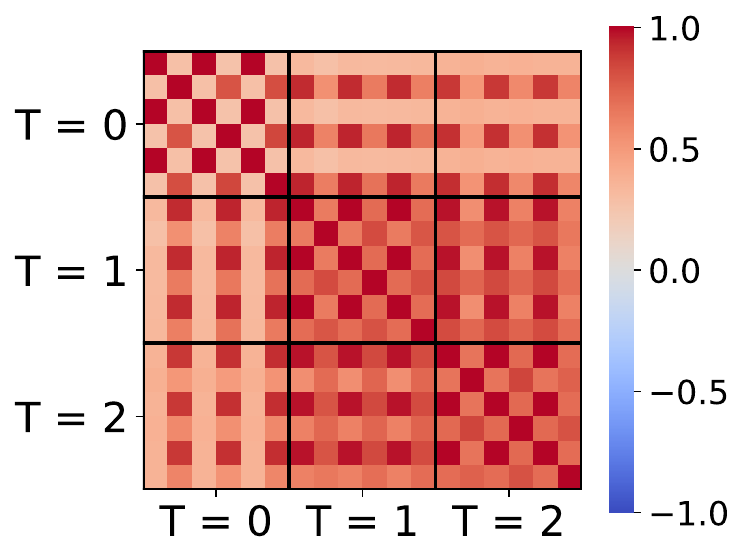}\label{fig:debate correlation}}
    \caption{Spatial and temporal correlations of three agents.}
    \label{fig:correlation}
\end{figure}

{To characterize stage-level spatial and temporal correlations, we model the prompt/completion length of each LLM stage and the duration of each non-LLM stage as random variables, and visualize their pairwise Pearson correlations for three widely used agents in Fig. \ref{fig:correlation}, where stages are grouped by temporal iteration $T$: diagonal blocks represent spatial correlations within the same iteration, and off-diagonal blocks capture temporal correlations across iterations. Self-Consistency shows near-perfect spatial correlation among prompt lengths since parallel samples share the same input prompt, while iterative agents such as Debate and ReAct exhibit structured temporal dependencies across successive rounds.}

\newpage

\section{Complete Algorithms}
\label{app:algorithms}
The following algorithms detail agent adaptation, agent scheduling, and model scheduling.

\begin{algorithm}[H]

\caption{Pareto-sampling-based Agent Adaptation}
\label{alg:agent_adaptation}
\begin{algorithmic}[1]
\REQUIRE Query $x$, quality predictor $P$, latency estimator $L$, model pool $\mathcal{M}$, agent pool $\mathcal{A}$, system snapshot $S$, sample budget $K$, quality target $q_\text{tar}$
\ENSURE Best agent-model candidate $c^\star$

\STATE $\mathcal{C} \gets \emptyset, \mathcal{E} \gets \emptyset$
\FORALL{$a \in \mathcal{A}$}
    \FORALL{$m \in \mathcal{M}$}
        \STATE $c \gets \textsc{UniformAssign}(a,m)$
        \STATE $c.qual \gets P(x,c)$, $c.lat \gets L(c,S)$
        \STATE $\mathcal{C} \gets \mathcal{C} \cup \{c\}$
    \ENDFOR
\ENDFOR

\FORALL{complex agent $a \in \mathcal{A}$}
    \STATE $\mathcal{F}_a \gets \textsc{ParetoFrontier}$ of uniform candidates of $a$
    \STATE $m_{\mathrm{lat}} \gets \textsc{LatencyAnchorModel}(\mathcal{F}_a, q_{\mathrm{tar}})$
    \STATE $m_{\mathrm{qual}} \gets \textsc{QualityAnchorModel}(\mathcal{F}_a, q_{\mathrm{tar}})$
    \IF{$m_{\mathrm{lat}}$\ exists\ and\ $m_{\mathrm{qual}}$\ exists}
        \STATE $\mathcal{E} \gets \mathcal{E} \cup \textsc{DeriveAreas}{(a, m_{\mathrm{lat}},
        m_{\mathrm{qual}}}, \mathcal{C})$
    \ENDIF
\ENDFOR

\STATE $\mathcal{C} \gets \mathcal{C} \cup \textsc{SampleAndEvaluate}(\mathcal{E}, K)$

\STATE $\mathcal{F} \gets \{c \in \mathcal{C} \mid c.qual \ge q_{\mathrm{tar}}\}$
\IF{$\mathcal{F} \neq \emptyset$}
    \STATE $c^\star \gets \arg\min_{c \in \mathcal{F}} c.lat$
\ELSE
    \STATE $c^\star \gets \arg\max_{c \in \mathcal{C}} c.qual$
\ENDIF

\STATE \textbf{return} $c^\star$
\end{algorithmic}
\end{algorithm}

\newpage

\begin{algorithm}[H]

\caption{Probabilistic Agent Scheduler}
\label{alg:prob_job_scheduler}
\begin{algorithmic}[1]
\STATE \textbf{Procedure} ScheduleAgents(\textbf{Active agents} $A$, \textbf{Idle batches} $\mathbf{B}=\{B_m\}_{m\in\mathcal{M}}$)
    \STATE $\mathcal{C} \gets \emptyset$
    \FORALL{$a \in A$ such that $a$ is unfinished}
        \STATE $z_a \gets \Call{NumNonresidentModels}{a}$
        \STATE $\mathcal{V}_a \gets \Call{ReadyStagesByCriticalPath}{a}$
        \STATE $\hat{R}_a \gets a.\Call{DBNEstimation}{} - \alpha W_a$
        \FORALL{$v \in \mathcal{V}_a$}
            \STATE $\mathcal{C} \gets \mathcal{C} \cup \{(a, z_a, \hat{R}_a, v)\}$
        \ENDFOR
    \ENDFOR

    \STATE Sort $\mathcal{C}$ by $(z_a, \hat{R}_a, v.CP\_rank)$ in ascending order
    \STATE $\mathcal{Q} \gets [\,]$

    \FORALL{$(a, z_a, \hat{R}_a, v) \in \mathcal{C}$}
        \STATE $m \gets v.model$
        \IF{$B_{m} > 0$}
            \STATE append $v$ to $\mathcal{Q}$
            \STATE $B_{m} \gets B_{m} - 1$
        \ENDIF
    \ENDFOR

    \STATE \textbf{return} $\mathcal{Q}$
\STATE \textbf{End procedure}

\end{algorithmic}
\end{algorithm}

\newpage

\begin{algorithm}[H]

\caption{Topology-aware Model Scheduling}
\label{alg:topo_model_scheduler}
\begin{algorithmic}[1]
\REQUIRE Active agents $A$, server state $\mathcal{S}$, model capacities $\{C_m\}_{m \in \mathcal{M}}$
\ENSURE GPU reallocation decisions

\STATE $\{D_m\} \gets \textsc{EstimateCurves}(A,\mathcal{S})$
\STATE $U_m[k] \gets D_m[k]-C_m$ for each model $m$
\STATE $\mathcal{M}^{-} \gets \textsc{SortUnderProvisioned}(\{m:\max U_m>0\})$
\STATE $\mathcal{M}^{+} \gets \textsc{SortOverProvisioned}(\{m:\max U_m<0\})$

\FORALL{$m^{-} \in \mathcal{M}^{-}$}
    \WHILE{$\max U_{m^{-}}>0$ and $\mathcal{M}^{+}\neq\emptyset$}
        \STATE $m^{+} \gets \textsc{BestDonor}(\mathcal{M}^{+})$
        \STATE $g^{\star} \gets \textsc{EarliestSwitchableGPU}(m^{+},\mathcal{S})$
        \IF{$g^{\star}=\emptyset$}
            \STATE remove $m^{+}$ from $\mathcal{M}^{+}$; \textbf{continue}
        \ENDIF
        \STATE \textsc{ReallocateGPU}$(g^{\star},m^{+}\rightarrow m^{-})$
        \STATE Update $U_{m^-},U_{m^+}$
    \ENDWHILE
\ENDFOR
\end{algorithmic}
\end{algorithm}

\end{document}